\documentstyle[aps,preprint,amssymb]{revtex} 
\begin{document}

\newcommand{\be}{\begin{equation}}
\newcommand{\bea}{\begin{eqnarray}}
\newcommand{\ee}{\end{equation}}
\newcommand{\eea}{\end{eqnarray}}
\newcommand{\ethbar}{\bar{\eth}}
\newcommand{\Lambdabar}{\bar{\Lambda}}
\newcommand{\p}{\partial} 
\newcommand{\zetabar}{\bar{\zeta}}
\newcommand{\omegabar}{\overline{\omega}}
\newcommand{\mbar}{\overline{m}}
\newcommand{\etabar}{\overline{\eta}}
\newcommand{\rhobar}{\overline{\rho}}
\newcommand{\s}{\sigma}
\newcommand{\sigmabar}{\overline{\sigma}}
\newcommand{\alphabar}{\overline{\alpha}}
\newcommand{\Xibar}{\overline{\Xi}}
\newcommand{\th}{\theta}
\newcommand{\Ld}{\Lambda}
\newcommand{\ld}{\lambda}
\newcommand{\tLd}{\tilde\Lambda}
\newcommand{\om}{\omega}
\newcommand{\ombar}{\bar{\omega}}
\newcommand{\Om}{\Omega}
\newcommand{\tOm}{\tilde\Omega}
\newcommand{\z}{\zeta}
\newcommand{\zbar}{\bar{\zeta}}
\newcommand{\Ldp}{\Lambda,_+}
\newcommand{\Ldm}{\Lambda,_-}
\newcommand{\Ldo}{\Lambda,_1}
\newcommand{\Ldn}{\Lambda,_0}
\newcommand{\varid}{\stackrel{\rm def}{=}}
\newcommand{\k}{\vec{k}}
\newcommand{\kcheck}{\check{k}}
\newcommand{\khat}{\hat{k}}
\newcommand{\mI}{($\tilde{m}_{I}$) }
\newcommand{\mII}{($\tilde{m}_{II}$) }
\newcommand{\g}{{\bf\mbox{g}}}
\newcommand{\f}{{\bf\mbox{f}}}
\newcommand{\scrip}{\cal I^+} 
\newcommand{\scrim}{\cal I^-}
\newcommand{\scri}{\cal I}
\newcommand{\ihat}{\hat{\imath}} \newcommand{\jhat}{\hat{\jmath}}
\newcommand{\Hspace}{${\cal H}$-space }
\newcommand{\lone}{\Lambda_1} \newcommand{\lonebar}{\Lambdabar_1}
\newcommand{\etatw}{\tilde \eta} \newcommand{\zetatw}{\tilde \zeta}

\title{Superselection Sectors in Asymptotic Quantization of Gravity}

\author{
	Alfredo E. Dominguez\thanks{e-mail: domingue@fis.uncor.edu}
\and
	Carlos N. Kozameh\thanks{e-mail: kozameh@fis.uncor.edu} 
       }

\address{FaMAF,
         Universidad Nacional de C\'ordoba,
         (5000) Cordoba, Argentina.
        }

\author{ 
	Malcolm Ludvigsen\thanks{e-mail: malud@mai.liu.se} 
       }

\address{Department of Mathematics
         Linkoping University
         S-581 83 Linkoping
         }
\maketitle 

\vspace{1cm}

\vspace{4cm}
\begin{abstract}
Using the continuity of the scalar $\Psi_2$ (the mass aspect) at null
infinity through $i_o$ we show that the space of radiative solutions of
general relativity can be thought of a fibered space where the value of
$\Psi_2$ at $i_o$ plays the role of the base space. We also show that
the restriction of the available symplectic form to each ``fiber'' is
degenerate. By finding the orbit manifold of this degenerate direction
we obtain the reduced phase space for the radiation data. This reduced
phase space posses a global structure, i.e., it does not distinguishes
between future or past null infinity. Thus, it can be used as the space
of quantum gravitons.  Moreover, a Hilbert space can be constructed on
each ``fiber'' if an appropriate definition of scalar product is
provided. Since there is no natural correspondence between the Hilbert
spaces of different foliations they define superselection sectors on
the space of asymptotic quantum states.

We discuss the physical relevance of the superselection sectors and
show that the analogous construction for linearized gravity yields
completely different results, thus emphasizing the need to use the full
nonlinearity of the theory even when discussing asymptotic
quantization.

\end{abstract}

\newpage

\section{Introduction}\label{}%

\vspace{.2 in}

The Null Surface Formalism (NSF) shows that General Relativity can be
viewed as a theory of hypersurfaces on a 4-dim manifold rather than a
field theory for a metric with lorentzian signature.  At a kinematic
level NSF shows that if two (complex) conditions are imposed on these
surfaces they become null hypersurfaces of a given metric. Field
equations equivalent to the vacuum Einstein equations determine the
dynamics of these characteristic hypersurfaces\cite{nullsurface}.

Within this formalism it is also possible to distinguish radiative
solutions of the vacuum equations. It can be shown that the Bondi free
data at $\scri$ enters as a source term in the NSF field equations for
null surfaces that are asymptotic planes at null infinity.  Thus,
for each Bondi data, the regular solutions to the field equations
represent global null surfaces of an asymptotically flat, vacuum
metric\cite{nullsurface}.

Recently, a paper extending the NSF to the quantum level was
presented\cite{fuzzy}. The starting point in that work is the
(classical) field equation that yields global null surfaces, i.e. null
surfaces associated with radiative solutions.  Adopting Ashtekar's
asymptotic quantization procedure\cite{aa:aq}, the Bondi free data of
the NSF equations is promoted to a quantum operator that obeys
commutation relations at null infinity. It then follows from the field
equations that the null surfaces themselves become operators that obey
non-trivial commutation relations\cite{fuzzy}. Furthermore, since it is
possible to identify points of the space-time as intersections of null
surfaces, it can also be shown that the space-time points themselves
become quantum operators.

There are however, technical difficulties in trying to construct a
physically relevant Hilbert and associated Fock space of incoming or
outgoing gravitons where these operators could act. As shown by
Ashtekar, it is possible to define a Hilbert space at null infinity but
doing so imposes severe restrictions on the free data of the associated
phase space since one leaves out almost all physically relevant spaces
\cite{aa:aq}.

In this paper we analyse again the phase space of asymptotic states. We
show that, using an exact conservation law, it is possible to foliate
this space in sectors that admit non-trivial scattering at a classical
level. Furthermore, we also show that the restriction of the symplectic
form to each foliation is degenerate and that a Hilbert space can be
constructed in each sector if one factors out the degenerate
direction.

In Section II we first review some results obtained in the context of
asymptotically flat space-times and present a theorem that is very
important for the main result of this work. 

In Section III we define the phase space of radiative modes and
introduce the notion of global structures on this space. We show that
\begin{enumerate}
\item the induced symplectic form on $\scri$, and 
\item a foliation on the phase space
adopting the value of $\psi_2$ at $i_o$ as the ``base'' space 
\end{enumerate}
are global structures.

In Section IV we show that the restricted symplectic form to each foliation is
degenerate.  We study the degeneracy direction and obtain the orbit
space associated with this direction. We introduce a complex structure on the tangent space to each fibre and construct, via
the symplectic form, a global, positive definite inner product.

Finally, in Section V we review the main results of this work, and
discuss possible generalizations. The application of this formalism to
the Maxwell field is given in the appendix.

\section{Radiative Spacetimes}\label{rad}

We define a radiative spacetime $M$ to be a solution of the empty
Einstein equations which is asymptotically flat at both future and past
null-infinity and suitably regular at spacelike and time-like
infinity.  We also require that it contains no horizons and is
topologically trivial in the sense that there exists a global
coordinate system.  This condition eliminates, for example, the
Schwarzschild solution.

A word of caution is appropriate here:  At the present moment no {\em
explicit} solution of Einstein's equations satisfying these conditions
has been found, and even the existence of such solutions has yet to be
rigourously demonstrated.  However, recent results, namely the Null
Surface Formalism, though not in its final form, does indicate that
radiative spacetimes do exist and that they can actually be constructed
from a single function, $\s$, which can be taken to be the asymptotic
shear on future null-infinity $\scrip$ (or $\scrim$).  By taking $\s$ to be
the asymptotic shear on $\scrip$ with a Bondi scaling, this construction
determines a radiative solution unique up to diffeomorphisms which
preserve the structure at null-infinity.

Another difficulty concerns the regularity conditions at space-like
infinity imposed in \cite{Xi}.  Although it would appear that these
conditions  are too restrictive for any physical spacetime to be
included in this class, it has been shown by M. Herberthson
\cite{Herberthson} that the conditions can be relaxed and still allow
sufficient regularity for the results of \cite{Xi} to be valid.  This
class includes all known asymptotically flat spacetimes.

The physical interpretation of a radiative spacetime is that of a
classical scattering event involving pure gravitational radiation, the
in-state being defined by $\s^-$ on $\scrim$ and the corresponding
out-state by $\s^+$ on $\scrip$.

Asymptotic flatness at null and space-like infinity implies the
existence of an extended conformally related space containing a null
cone representing points at infinity.  Its vertex $i^o$ represents
space-like infinity and its future and past parts, $\scrip$ and $\scrim$,
past and future null infinity.  Except for flat spacetime, the point
$i^o$ is not smoothly attached but has a direction dependent
structure.  This, however, allows the introduction of a stereographic
coordinate function $\zeta$ on $I=I^+\cup I^-$ which is constant along
the null generators of $I$ and continuous through $i^o$.  This function
effectively provides a one-to-one correspondence between the generators
of $\scrip$ and $\scrim$.

In terms of this type of scaling, $\scrip$ is diverging and $\scrim$ is 
converging.  A Bondi scaling, on
the other hand, makes $\scrip$ and $\scrim$ divergence free and thus $i^o$ 
becomes an infinitely removed
point.  Given a stereographic coordinate function $\zeta$ on $I$ we 
introduce a particular Bondi
conformal factor $\Omega$ such that any space-like slice of $I$ has an 
induced metric of the form
$$ds^2=\frac{d\zeta d\bar{\zeta}}{P^2}$$
where $2P=1+\zeta\bar{\zeta}$.  This is possible because all space-like
slices are isometric by virtue of the divergence-free condition.  On
$I$ the vector $n_a=-\nabla_a\Omega$ is null, non-zero, and points
along the generators.  On $\scrip$ it is future pointing and on $\scrim$ it
is past pointing.

  We now introduce a Bondi parameter function $u$ on $I$ satisfying 
$n^a\nabla_a u=1$
which determines $u$ up to
$$u\to u+\gamma$$
where $\gamma$ is constant along the generators.  Since $I$ consists of 
two disconnected
components, $\gamma=(\gamma^+,\gamma^-)$ where $\gamma^{\pm}$ are the 
restrictions of $\gamma$ to
$I^{\pm}$.  These two functions may be chosen independently.  Note that 
on both $\scrip$ and $\scrim$
$u=-\infty$ represents space-like infinity.  We now have a global 
coordinate system
$(u,\zeta,\bar{\zeta})$ which labels points on both $\scrip$ and $\scrim$.

Given a Bondi parameter $u$, we complete $n^a$ to form a null-tetrad 
$(n^a,m^a,\bar{m}^a,l^a)$
 on $I$ by demanding that $m^a$ is tangent the $u=$ constant slices.  
This determines $l^a$
uniquely on $I$ (it points out of $I$ and is future-pointing on $\scrip$ and 
past-pointing on $\scrim$)
but $m^a$ only up to 
$$m^a\to e^{i\lambda}m^a.$$ 
A tetrad dependent function $\eta$ on $I$ which transforms according to 
$$\eta\to e^{is\lambda}\eta$$
is said to have spin-weight $s$.   Though not strictly necessary, it is 
convenient to fix $m^a$ by
demanding $m^a\nabla_a\bar{\zeta}=0$.

Two important tetrad dependent functions on $I$ are the shear  $\s$ and 
the mass aspect
$\psi_2$ which are defined by
\begin{eqnarray}%
\s(u,\zeta,\bar{\zeta})     &=& m^am^b\nabla_a l_b \\
\psi_2(u,\zeta,\bar{\zeta}) &=& \Omega^{-1} C_{abcd} n^a \bar{m}^b l^c m^d,
\end{eqnarray}
Note that $\s$ and $\psi_2$ have spin-weights 2 and 0 respectively.  We
demand smoothness in $u$ and regularity in the angular coordinates
$(\zeta,\bar{\zeta})$ in the sense that $\s$ and $\psi_2$ are
expandable in terms of the appropriate spin-weighted spherical
harmonics.  We also demand that the limits $\lim_{u\to\pm\infty}\s$ and
$\lim_{u\to\pm\infty}\psi_2$ exist.   Since $I$ consists of two
disconnected components, we have $\s=(\s^+,\s^-)$ where $\s^{\pm}$ are
the restrictions of $\s$ to $I^{\pm}$, and similarly for $\psi_2$.

By regularity  at time-like infinity we mean
\begin{equation}\label{1.1}%
\lim_{u\to\infty}\psi_2=0 \quad (\lim_{u\to\infty}\psi^{\pm}_2=0)
\end{equation}%
By an extension of the positive mass theorem \cite{LudVic} this implies 
that the initial data on an
asymptotically null, space-like hypersurface, which intersects $I$ at 
$u=$ constant, becomes flat
for $u\to\infty$.  Since we do not have flatness at space-like infinity
\begin{equation}\label{1.2}%
\lim_{u\to - \infty}\psi_2\ne 0 \quad (\lim_{u\to -\infty}\psi^{\pm}_2 
\ne 0).
\end{equation}%
However, the results of \cite{Xi} imply
\begin{equation}\label{1.3}%
\lim_{u\to - \infty}\psi^+_2=\lim_{u\to - 
\infty}\psi^-_2=\chi(\zeta,\bar{\zeta}).
\end{equation}
This equation is important since it provides a link between $\scrip$ and 
$\scrim$. 

On $I$ we have the spin-coefficient relations \cite{NP68}:
\begin{equation}\label{1.4}%
\dot{\psi_2}=-\eth^2\dot{\bar{\s}}-\s\ddot{\bar{\s}}
\end{equation}%
and
\begin{equation}\label{1.5}%
\psi_2-\bar{\psi_2}=\bar{\s}\dot{\s}+\bar{\eth}^2\s -c.c
\end{equation}
Using eqs (\ref{1.5}) and (\ref{1.1}) we have
$$\lim_{u\to\infty}(\bar{\eth}^2\s-\eth^2\bar{\s})=0$$
which implies the existence of a Bondi frame (frames) such that
\begin{equation}\label{1.6}%
\lim_{u\to\infty}\s=0 \quad (\lim_{u\to\infty}\s^{\pm}=0) 
\end{equation}%

Using equations (\ref{1.4}), (\ref{1.6}) and (\ref{1.3}) we now have
\begin{equation}\label{1.7}%
\chi=\bar{\eth}^2\s^{\pm}_o+\int^\infty_{-\infty}\dot{\s}^{\pm}\dot{\bar{\s}}^{\pm}du
\end{equation}%
where $\s^{\pm}_o=\lim_{u\to -\infty}\s^{\pm}$.  This provides a very
important relation between in-states described by $\s^-$ and out-states
described by $\s^+$. Classical scattering sends  in-states to out-states associated with the same function $\chi$. We exploit this feature in the next section to foliate the reduced phase space of radiative solutions.

Given a real function $\kappa(\zeta,\bar{\zeta})$ equation (\ref{1.7}) gives
\begin{equation}\label{1.8}%
\chi_\kappa=\s_\kappa^{\pm}+P_\kappa^{\pm}
\end{equation}%
where
\begin{equation}\label{1.9}%
\s_\kappa^\pm=\int \kappa \eth^2 \s_o^\pm dS
\end{equation}%
and %
\begin{equation}\label{1.10}%
P_\kappa^{\pm}=\int_{I^\pm}\kappa \dot{\s}\dot{\bar{\s}}dI.
\end{equation}%
Here $dS$ is the area element on the $u=$ constant slices and $dI=dSdu$.
$P_\kappa^{\pm}$ is the total flux of super momentum through $\cal I^\pm$.  If 
$\kappa$ satisfies
\begin{equation}\label{1.11}%
\eth^2\kappa=0
\end{equation}%
$P_\kappa^{\pm}$ reduces to
the total flux of Bondi momentum (in the direction defined by $\kappa$)
through $I^\pm$ and, by integrating by parts, (\ref{1.8}) gives the
asymptotic conservation law

\begin{equation}\label{1.12}%
P_\kappa^+=P_\kappa^-.
\end{equation}

It is interesting to consider the outline of another proof of this
conservation law which brings out the necessity of condition
(\ref{1.11}) in a more geometrical way.  By using a variation of the
Ludvigsen-Vicker's proof of the positivity of Bondi mass \cite{LudVic},
it can be shown that there exists a spinor field $\kappa_A$ in the
interior of a radiative spacetime which induces a super-translation
$\kappa$ on $I$ and an exact 3-form $j_\kappa$ in the interior such
that $$P_\kappa^\pm=\int_{I^\pm}j_\kappa.$$ This does not lead directly
to a conservation law $P_\kappa^+=P_\kappa^-$ since $j_\kappa$ is in
general  singular at $i^o$.  However, in the special case where
$\kappa$ is a translation in that it satisfies (\ref{1.11}),
$j_\kappa$ becomes sufficiently regular at $i^o$ for Stokes' theorem to
hold and this leads to the conservation law $P_\kappa^+=P_\kappa^-$ for
the total flux of Bondi momentum.   The only way to obtain a
conservation law for super momentum is to assume $j_\kappa=0$ at $i^o$
but this is true only for flat space.  Nevertheless, an interesting
result for super momentum can be obtained for first-order perturbations
of a radiative spacetime.  In this case a perturbation can be chosen
which does not affect space-like infinity in the sense that
$\delta\chi=0$.  This gives $\delta j_\kappa=0$ at $i^o$ and a direct
application of Stokes' theorem gives the perturbative conservation law
for super momentum:
\begin{equation}\label{1.13}%
\delta P_\kappa^+=\delta P_\kappa^-.
\end{equation}%
Combining this with equation (\ref{1.8}) we obtain the following 
important result:

\vspace{.5cm}

\noindent {\bf Lemma:} {\it Given a perturbation such that $\delta\chi=0$ and $\delta \s^-_o=0$, then the same perturbation when propagated to $\scrip$ gives
$\delta\s^+_o=0$.}

\section{Phase Space of Radiative States}\label{phase}%

We define the non-reduced phase space of radiative states to be the set
$\Sigma$ of all pairs $\s=(\s^+,\s^-)$ where $\s^+$ and $\s^-$ are
shear functions `joined' by some radiative spacetime in the sense of
the previous section.  The reduced phase space can be obtained by
factoring out the physically irrelevant structure provided by the
choice of the coordinate functions $u$ and $\zeta$.  This
gives a space where each point corresponds to a radiative spacetime
unique up to diffeomorphisms.  Since $\s$  transforms quite simply
under a change of Bondi frame, such a reduction can easily be obtained,
but for the sake of simplicity we shall content ourselves with the
non-reduced phase space $\Sigma$.

The Bondi shears $\s^-$ and $\s^+$ determine an in-state and out-state
respectively. By construction they satisfy

$$\lim_{u\to\infty}\s^{\pm}=0,$$ 
together with the smoothness and regularity properties stated in the
previous section.  A given in-state $\s^-$ determines a corresponding
out-state $\s^+$ up to a BMS translation.  Apart from this, all we know
about the relation between $\s^-$ and $\s^+$ is that given by equation
(\ref{1.7}).  This provides a geometrically determined foliation on
$\Sigma$ where two points lie in the same leaf if they determine the
same function $\chi(\zeta,\bar{\zeta})$. We thus have as many
equivalence classes on ${\Sigma}$ as there are regular complex
functions on $S^2$.

A natural question to ask is whether $\s^-$ may be chosen freely
subject to the conditions already stated.  The Null Surface Formalism
shows that this is actually the case:  given any function
$\s^-(u,\zeta,\bar{\zeta})$ satisfying these conditions, a radiative
spacetime can, in principle, be constructed together with a Bondi frame
$(u,\zeta,\bar{\zeta})$ such that $\s^-(u,\zeta,\bar{\zeta})$ is its
past shear.

By time-reversal symmetry, we see that $(\s^+,\s^-)\in \Sigma$ implies
that $(\s^-,\s^+)\in \Sigma$.  A geometrical structure on $\Sigma$
which does not depend on a preference between $\s^+$ and $\s^-$ will be
said to be global.  Global structures are particularly important as
regards full quantization as apposed to asymptotic quantization.  The
foliation defined by the function $\chi$ is, for example, a global
structure in this sense.  The subspace of $\Sigma$ defined by
$$\lim_{u\to -\infty}\s^-=\s^-_o=0$$ is not, however, a global
structure since this condition does not imply 
$$\lim_{u\to -\infty}\s^+=\s^+_o=0.$$

To obtain another type of global structure, namely a symplectic form,
we must consider the space $T_\s(\Sigma)$ of tangent vectors a some
point $\s\in \Sigma$.  A tangent vector at $\s$ (corresponding to a
spacetime $M$) may be viewed as a perturbation
$\delta\s=(\delta\s^+,\delta\s^-)$ where $\s+\delta\s^+$ and
$\s+\delta\s^-$ are joined by a perturbed space $M+\delta M$.  Starting
from the standard symplectic form defined on a Cauchy surface and using
a conserved current in the interior of the spacetime $M$ it is possible
to show that\cite{AshMagnon}
\begin{eqnarray}
\Omega (\delta \s_1,\delta\s_2)&=&
\int_{I^-}{\rm d}^{3}I\;[{\delta\sigma}^-_1 {\delta{\dot{\bar\sigma}}}^-_2
-{\delta{\dot{\bar\sigma}}}^-_1{\delta\sigma}^-_2] + \; c.c.\nonumber\\ 
&=&\int_{I^+}{\rm
d}^{3}I\;[{\delta\sigma}^+_1 {\delta{\dot{\bar\sigma}}}^+_2
-{\delta{\dot{\bar\sigma}}}^+_1{\delta\sigma}^+_2] + \; c.c., \label{sym}
\end{eqnarray}
This defines a global, non-degenerate, symplectic form on $\Sigma$.
 
Our phase space $\Sigma$ is thus a foliated, symplectic space.  The
symplectic form and foliation are preserved under reduction, where a
point in the reduced space is now an equivalence class $\hat{\s}$ of
related $\s$'s and the foliation is determined by an equivalence class
$\hat{\chi}$ of related $\chi$'s.

\section{Tangent Vector Spaces}\label{tan}%

We shall now restrict attention to tangent vectors $\delta\s$ such that 
\begin{equation}\label{2.1}%
\lim_{u\to -\infty}\delta\s^-=0.
\end{equation}%
We emphasise that this is not a global geometric condition since it does 
not imply
\begin{equation}\label{2.2}%
\lim_{u\to -\infty}\delta\s^+=0.
\end{equation}%
but it is necessary for the construction of a hilbert space, which is
normally the first step towards quantization.  Our basic idea is to
find a subspace of $T_\s(\Sigma)$ (subject to this condition) which is
global in that both (\ref{2.1}) and (\ref{2.2}) are satisfied, and
which admits a natural hilbert-space structure.

Let us first see how a natural, but non-global, hilbert-space structure
can be defined on $T_\s(\Sigma)$ when subject to condition
(\ref{2.1}).  By construction we have $\lim_{u\to\infty}\delta\s^-=0$
and thus $\delta\s^-$ tends to zero at both ends of $\scrim$.  This
means that $\delta\s^-$ admits a Fourier decomposition with respect to
$u$ and, by means of this, we can find the positive and negative
frequency parts, $\delta\s_{pos}^-$ and $\delta\s_{neg}^-$, of
$\delta\s^-$.  The complex structure $J$ corresponding to this
decomposition is defined by
\begin{equation}\label{2.3}%
J\delta\s=((J\delta\s)^+,(J\delta\s)^-)
\end{equation}%
where
\begin{equation}\label{2.4}%
(J\delta\s)^-=i(\delta\s_{pos}^- - \delta\s_{neg}^-)
\end{equation}%
With respect to this complex structure, multiplication by a complex 
number $z=x+iy$ is defined by
$$z\delta\s=x\delta\s+yJ\delta\s.$$
It can easily be seen that $J$ is compatible with $\Omega$ in that
$$\Omega(J\delta\s_1,J\delta\s_2)=\Omega(\delta\s_1,\delta\s_2)$$
and positive in that
$$\Omega(\delta\s,J\delta\s)>0$$
for non-trivial $\delta\s$.  From these results we see that
\begin{equation}\label{2.5}%
\langle \delta\s_1,\delta\s_2
\rangle=\Omega(\delta\s_1,J\delta\s_2)+i\Omega(\delta\s_1,\delta\s_2)
\end{equation}%
is a positive-definite, non-degenerate hermitian product, i.e.,

\begin{eqnarray}%
\langle \delta\s_1,\delta\s_2\rangle&=&
\overline{\langle\delta\sigma_2,\delta\sigma_1\rangle}\\
\langle\delta\s,z_1\delta\s_1+z_2\delta\s_2\rangle&=&
z_1\langle\delta\s,\delta\s_1\rangle+z_2\langle\delta\s,\delta\s_2\rangle\\
\langle\delta\s,\delta\s\rangle &>& 0 \mbox{ if } 
\delta\hat{\sigma}\ne 0
\end{eqnarray}%

The tangent space $T_\s(\Sigma)$ thus has a natural hilbert space
structure.  This structure is also preserved under coordinate reduction
but, as we have already pointed out, it is not global in that it is
defined with respect to $\scrim$.

It is interesting to note at this point that $T_\s(\Sigma)$ possesses an
even more natural complex structure defined simply by $J=i$.  This is a 
rather curious fact
because the basic elements we are dealing with are radiative spacetimes 
which are essentially
real objects, and yet we end up with a complex vector space, in fact a 
complex vector space with two
complex structures.

By the final result in section (\ref{rad}) we see that the subspace $H$
of $T_\s(\Sigma)$ consisting of vectors which satisfy $\delta\chi=0$,
and which therefore lie in the leaf containing $\s$, is global in that
$\delta\s_o^-=0$ implies $\delta\s_o^+=0$.
From equation (\ref{1.7}) we see that an element $\delta\s$ of
$T_\s(\Sigma)$  is contained in $H$ iff

\begin{equation}\label{2.6}%
\alpha(\delta\s)\equiv 
\int_{-\infty}^{\infty}(\delta\s\ddot{\bar{\s}}+\delta\bar{\s}\ddot{\s})du=0.
\end{equation}
The space $H$ contains, in turn, a preferred subspace $K$ consisting of 
vectors $\delta\s_k$ such that
$\delta\s_k^-=\kappa\dot{\s}^-$ where $\kappa$ is real and 
$\dot{\kappa}=0$.  Using (\ref{sym}) it
can easily be checked that condition (\ref{2.6}) is equivalent to
\begin{equation}\label{2.7}%
H=\{ \delta\s:\Omega(\delta\s,\delta\s_k)=0,\; \forall \, \delta\s_k\in K \}.
\end{equation}%
This defines $H$ in terms of $K$, thus showing that $K$ is global in
spite of the fact that it is defined with respect to $\scrim$, and also
shows that $K$ contains all directions of degeneracy of $\Omega$ when
restricted to $H$.  [A vector $\delta\s_k$ is a direction of degeneracy
of $\Omega$ restricted to $H$ if $\Omega(\delta\s,\delta\s_k)=0$ for
all $\delta\s\in H$.]

In order to obtain a space with a non-degenerate symplectic product we
factor out the directions of degeneracy and consider the space $\hat{H}$ of equivalence classes where $\delta\s$ and $\delta\s'$ belong to the
same equivalence class $\delta\hat{\s}$ if $\delta\s-\delta\s'\in K$.

The symplectic product defined by
$$\hat{\Omega}(\delta\hat{\s}_1,\delta\hat{\s}_2) = \Omega(\delta\s_1,\delta\s_2)$$
is non-degenerate on $\hat{H}$.

To understand the physical meaning of this reduced phase space we study the integral lines of $\delta\sigma_k$, which are obtained by introducing a
parameter $s$ and solving

\begin{equation}
\frac{d \sigma_k}{d s} = \kappa \frac{d \sigma}{d u},
\end{equation}

whose solution is given by

\begin{equation}
\sigma_s(u,\zeta,\bar\zeta) = \sigma(u + \kappa s,\zeta,\bar\zeta).
\label{10}
\end{equation}

Thus, for each value of $(\zeta,\bar\zeta)$ we have to factor out Bondi
data that are supertranslations of a given function $\sigma(u, \zeta, \bar\zeta)$ by an arbitrary distance along the timelike
direction $u$.

The manifold of orbits of $\delta\sigma_k$ is the reduced phase space
$\hat{\Sigma}$ and $\hat{H}$ its associated tangent space. 

The complex structure $J$ does not induce a complex structure on $H$
since $\delta\s\in H$ does not imply $J\delta\s\in H$, or,
equivalently, $\alpha(\delta\s)=0$ does not imply
$\alpha(J\delta\s)=0$.  In particular, we have
\begin{equation}\label{2.8}%
\alpha(J\delta\s_o) \ne 0
\end{equation}%
for all non-trivial $\delta\s_o\in K$.  Fortunately, we are not so much
interested in $H$ as its reduced version $\hat{H}$ consisting of
equivalence classes of $H$.  Using (\ref{2.8}), we see that each such
equivalence $\delta\hat{\s}$ class contains a unique representative
$\delta\s'$ such that $\alpha(J\delta\s')=0$ and hence $J\delta\s'\in
H$.  Since the space of all such vectors is clearly equivalent to
$\hat{H}$ we see that $J$ induces a complex structure $\hat{J}$ on
$\hat{H}$.  It is easily checked that $\hat{J}$ is compatible with
$\hat{\Omega}$ and positive.  It thus induces a hibert space structure
on $\hat{H}$.

The observant reader will have noticed that a similar construction
based on $\scrip$ rather than $\scrim$ leads to another compatible,
positive, complex structure $\hat{J}'$ on $(\hat{H},\, \hat{\Omega})$.
However, since $\hat{H}$ and $\hat{\Omega}$ are globally defined in the
sense that their definition is independent of the choice between
$\scrip$ and $\scrim$, it seems reasonable to conjecture that
$\hat{J}=\hat{J}'$.  Whether or not this is true, we have at least
shown the {\em existence} of a compatible, positive, complex structure
on  $(\hat{H},\, \hat{\Omega})$.

\section{Summary and Conclusions}\label{}%

We have shown that the phase space of the radiative degrees of freedom
of General Relativity can be foliated using the continuity of the mass
aspect ($\psi_2$) through $i_o$. It was also shown that this foliation
is a well defined global structure at $\scri$

We have then shown that the induced symplectic form on each foliation
is degenerate along a direction that represents a translation of the
free data $\sigma$ along each generator of $\scri$.  By factoring out
this degenerate direction one obtains the restricted phase-space
$\hat{\Sigma}$ associated with each foliation that also has a global
meaning

Introducing a complex structrure $J$ one then defines an inner product
on $\hat{H}$ that has a finite norm. This 1-graviton Hilbert space so
constructed is the building block for the Fock space associated with
each foliation.  Since there is no natural relation between Fock spaces
for different ``fibres'' this quantization procedure defines
superselection sectors on the full phase space.

As we will show in the appendix, if we repeat the same construction for
free Maxwell fields in flat space-time one can also find foliations
associated with the continuity of a Maxwell scalar ($\phi_1$) through
$i_o$. However, the induced symplectic form on each foliation is now
non-degenerate and thus, the procedure to construct a Hilbert space
follows a different approach.

The differences between the Maxwell case and General Relativity has
profound implications. If we had linearized the gravitational field
equations to construct the asymptotic phase space we would have
followed a similar approach to the Maxwell case and we would have
missed the fact that the induced symplectic form on each foliation for
full gravity is degenerate. The lesson being learned here is that the
superselection sectors for linearized gravity are manifestly different
from the ones that arise in the full theory.

At this time one could ask two questions:

\begin{enumerate}
\item Is this superselection rule physically significant?
\item Does the quantum S-matrix preserve the superselection sectors?
\end{enumerate}

In Maxwell theory each superselection sector is associated with
quantization of the radiation field in the presence of a classical
distribution of electromagnetic charges. This meaning extends to QED if
we replace the classical distribution of charges with the density
distribution of the Dirac field.

Likewise, in GR we associate superselection sectors with quantization
of the fluctuations of a classical distribution of mass. It is not
clear if a full quantum theory of gravity will admit superselection
sectors, though the mass operator should be a conserved quantity.

Although a quantum theory is needed to construct the S-matrix and thus
answer the second question, the answer is positive at a classical level
since we have shown that these foliations are global structures. Thus,
classical scattering preserves each foliation and the pushforward map
sends tangent vectors to a foliation at $\scrim$ to tangent vectors to
the corresponding foliation at $\scrip$. If the quantum S-matrix takes
a vector belonging to a particular foliation at $\scrim$ and produces a
vector that is not tangent to the same foliation at $\scrip$ then the
S-matrix will fail to be unitary since in general a vector at $\scrip$
that is not tangent to the foliation will not have a finite norm. To
make more assertive claims however, one needs the full quantum dynamical
evolution.

The idea for future work is to use this kinematic quantization with the
quantum NSF. As mentioned in the introduction, this quantization is
done at the operator level and we are here introducing the appropriate
Fock space where these operators act. The null cone quantization then
gives the dynamical evolution of these operators and, by properly
taking limits, one can construct the S-matrix of the theory.

\vspace{1cm}

\noindent{\bf Acknowledgments}:
E.D and C.K. wish to thank Abhay Ashtekar and Oscar Reula whose comments
and suggestions led to important  improvements in the
completeness and presentation of this work. \\
This research has been partially supported by CONICET and CONICOR.

\section*{Appendix A: Superselection sectors in Maxwell Theory}

\renewcommand{\theequation}{A.\arabic{equation}} \setcounter{equation}{0}

In this appendix we use the continuity of the scalar $\phi_1$ through
$i_o$ to foliate the solution space of radiative solutions to the
source free Maxwell's equations in Minkowski space. We show that the
restricted symplectic form to each foliation is non-degenerate and that
it yields a finite norm.

Since we work in the null tetrad formalism, instead of the Maxwell
field $F_{ab}$ we use the (complex) scalars

\begin{eqnarray*}
& & \phi_0=F_{ab}l^am^b\\
& & \phi_1=\frac{1}{2}F_{ab}(l^an^b+\bar m^am^b)\\
& & \phi_2=F_{ab}\bar m^an^b,\\
\end{eqnarray*}
where $(l^a, m^a,\bar m^a,  n^a)$ is a null tetrad adapted to the geometry
of compactified Minkowski space with null boundary $\scri$.

Assuming the source free Maxwell field has finite energy one can show that
the restriction of the scalar $\phi_1$ to $\scri$ satisfies
\begin{eqnarray}
& & \lim_{x \to {\cal i}^+}\phi_1(x)=0,\quad x \in {\cal I}^+\nonumber\\
& & \lim_{x \to {\cal i}^-}\phi_1(x)=0,\quad x \in {\cal I}^-.
\end{eqnarray}

Furthermore, a linearization of the approach presented in \cite{Xi} shows that $\phi_1$ is continuous through $i_o$, i.e.:

\begin{equation}
\lim_{x \to i_o}\phi_1(x)=\lim_{x' \to i_o}\phi_1(x'),\quad x \in {\cal
I^-}\quad {x'}\in {\cal I}^+ \label{A2}
\end{equation}

The idea is to use this continuity to foliate the solution space of
radiative solutions.

Since we are following a similar approach to the gravitational case we will be interested in global structures. We
construct the phase space ${\cal A}$ for Maxwell theory using the two
degrees of freedom of the restricted maxwell potential $A_a$ to
$\scri$. It can be shown that the complex function

\begin{equation}
A_\pm \equiv \lim_{x \to \scri^\pm}A_a m^a
\end{equation}
(with $A_a$ a potential in the gauge $A_a n^a = 0$ at $\scri^\pm$) captures
the two degrees of freedom of the radiative solutions. As before, for
simplicity we work on $\scrim$ and we drop the superscript on the scalars.

We define then the non-reduced phase space of radiation fields to be the set 
$\cal A$ of all pairs
$A=(A^+,A^-)$ where
$A^+$ and $A^-$ are  ``joined'' by a radiative solution of Maxwell's equations.  

Using the relationship between the field and potential one can show that

$$\phi_1|_{\cal I}=\eth\bar{A}.$$
Furthermore, since the kernel of the $\eth$ operator acting on s.w. 1
functions vanishes, there is a one to one correspondence between $\phi_1$
and
$A$ and thus, it follows from (\ref{A2}) that

\begin{equation}
\lim_{x \to i_o}A(x)=\lim_{x' \to i_o}A(x') = \Xi(\zeta, \bar{\zeta}),\quad
x , \quad {x'}\in {\cal I}^- \label{A3}
\end{equation}

As before, we have as many equivalence classes on ${\cal A}$ as there
are regular complex functions on $S^2$. This equivalence relation
introduces a foliation on ${\cal A}$. All $A$'s belonging to a
foliation are labelled by the function $\Xi(\zeta, \bar\zeta)$.

We denote by $T_A(\cal A)$ the tangent space at a point $A$ and by 
$\delta A=(\delta A^+,\delta A^-)$ a tangent vector on this space. It can be easily shown that from (\ref{A3}) and the linearity of Maxwell's equations, any tangent vector $\delta A$ is global. However, since our goal is to construct a Hilbert space on $T_A(\cal A)$, we will only consider vectors such that $\delta \Xi = 0$ (only these vectors admit a Fourier decomposition along the u direction). On those vectors $\delta\hat{A}$ we introduce the following complex structure

\begin{equation}\label{A.4'}%
J\delta \hat{A}=((J\delta \hat{A})^+,(J\delta \hat{A})^-)
\end{equation}%
where
\begin{equation}\label{A.5'}%
(J\delta \hat{A})^-=i(\delta \hat{A}_{pos}^- - \delta \hat{A}_{neg}^-)
\end{equation}%

We recall that the symplectic form defined in the
canonical formalism induces a global, non-degenerate form on $\scri$ given by.

\begin{equation}
\Omega (\delta A^1,\delta A^2) = \int {\rm d}^3 I\;[ \delta A^1\;\delta\dot
{\bar {A}^2} -\delta\dot {A}^1\;\delta {\bar A^2} ]+\;c.c,
\label{A4}
\end{equation}
where $\delta A^1$ and $\delta A^2$ are tangent vectors in the phase space.
The idea now is to restrict the (weakly) non-degenerate symplectic form
to each foliation, i.e.,  we evaluate (\ref{A4}) on
vectors $\delta\hat{A}$. It follows from
(\ref{A3}) that these vectors satisfy

\begin{equation}
\lim_{x \to i_o}\delta\hat{A}(x) = 0.
\label{A5}
\end{equation}

We want to show that this restricted form is non-degenerate. Assume
there exists a $\delta\hat{A}_o$ that satisfies (\ref{A5}) and

\begin{equation}
\Omega (\delta \hat{A}_o,\hat{A}) = 0
\end{equation}
for all $\delta\hat{A}$'s that belong to $T_{A, \Xi}(\hat{\cal A})$.
Integrating by parts this equation it then follows that

\begin{equation}
\delta\dot{\hat{A}_o} = 0,
\end{equation}
or
\begin{equation}
\delta\hat{A}_o = f(\zeta,\bar{\zeta}),
\end{equation}
which contradicts (\ref{A5}).

As before $J$ is compatible with $\Omega$ 
$$\Omega(J\delta A_1,J\delta A_2)=\Omega(\delta A_1,\delta A_2)$$
and is positive for non-trivial $\delta\s$.  We thus define the following 
positive-definite, non-degenerate hermitian product

\begin{equation}\label{2.5}%
\langle \delta A_1,\delta A_2
\rangle=\Omega(\delta A_1,J\delta A_2)+i\Omega(\delta A_1,\delta A_2)
\end{equation}%

It is worth mentioning that vectors $\delta A$'s that are not tangent
to a foliation, i.e., that do not satisfy (\ref{A5}) will have infinite
norm and thus will not belong to the Hilbert space associated with $\cal A$.
These states belong to the infrared sector and are fully discussed in \cite{aa:aq}.

On the other hand, all vectors $\delta \hat{A}$ tangent to the foliations
have finite norm and one thus defines a Hilbert space for each foliation.
Since there is no natural connection between the different Hilbert spaces
so constructed we call them superselection sectors.

What is the physical meaning of the superselection sectors?

We first observe that $\phi_1$ yields the charge aspect of Maxwell
fields. If we  allow for the presence of bounded sources then each
superselection sector yields the quantized radiation fields associated
with a particular charge configuration. This relationship can be
extended to full QED by constructing a fibre bundle where a point on
the base space corresponds to a Dirac state and the fibre above this
point is the corresponding superselection sector.

\end{document}